\documentclass[journal]{IEEEtran}
\usepackage{amsmath,amsfonts}
\usepackage{algorithmic}
\usepackage{algorithm}
\usepackage{array}
\usepackage[caption=false,font=normalsize,labelfont=sf,textfont=sf]{subfig}
\usepackage{textcomp}
\usepackage{stfloats}
\usepackage{url}
\usepackage{verbatim}
\usepackage{graphicx}
\usepackage{cite}
\usepackage{arydshln}
\usepackage{booktabs} 
\usepackage{pdfpages}
\usepackage{tabularx} 
\usepackage{multirow}
\usepackage{soul}
\usepackage{cancel}
\usepackage[normalem]{ulem}
\usepackage{xcolor}

\begin{document}

\title{Bi-modal Prediction and Transformation Coding for Compressing Complex Human Dynamics}

\author{Huong Hoang, Keito Suzuki, Truong Nguyen, \textit{Fellow}, IEEE, and Pamela Cosman, \textit{Fellow}, IEEE \\
Department of Electrical and Computer Engineering, University of California, San Diego, CA, USA
\thanks{This work was supported by the Center for Wireless Communications, University of California, San Diego, and by the National Science Foundation under Grant DUE-1928604. }}



\maketitle

\begin{abstract}
For dynamic human motion sequences, the original KeyNode-Driven codec often struggles to retain compression efficiency when confronted with rapid movements or strong non-rigid deformations. This paper proposes a novel Bi-modal coding framework that enhances the flexibility of motion representation by integrating semantic segmentation and region-specific transformation modeling. The rigid transformation model (rotation \& translation) is extended with a hybrid scheme that selectively applies affine transformations—rotation, translation, scaling, and shearing—only to deformation-rich regions (e.g., the torso, where loose clothing induces high variability), while retaining rigid models elsewhere. The affine model is decomposed into minimal parameter sets for efficient coding and combined through a component selection strategy guided by a Lagrangian Rate-Distortion optimization. The results show that the Bi-modal method achieves more accurate mesh deformation, especially in sequences involving complex non-rigid motion, without compromising compression efficiency in simpler regions, with an average bit-rate saving of 33.81\% compared to the baseline.
\end{abstract}


\section{Introduction}
The efficient compression of real-world scanned 3D human dynamic meshes presents a significant challenge at the intersection of geometry processing, data transmission, and immersive technologies. As applications such as telepresence, virtual and augmented reality, and 3D streaming become increasingly widespread, the ability to efficiently represent and transmit high-fidelity human motion data is crucial for enabling realistic and interactive experiences. However, scanned human meshes present complexities such as variable topology, scanning artifacts, and an interplay of rigid and non-rigid motion components. To address these challenges, various compression frameworks have emerged, notably those that leverage embedded key nodes. They typically model the temporal evolution of each vertex as a distance-weighted combination of transformations derived from a sparse set of control nodes, achieving data reduction by transmitting only the key nodes' transformations.

Despite their advantages, a significant limitation in some existing KeyNode-driven methods, particularly noticeable in regions undergoing substantial deformation, arises from their reliance on rigid transformations (combinations of rotation and translation). 
While effective for rigid or semi-rigid body prediction, this approach struggles with highly non-rigid deformations such as loose clothing, which can exhibit large and even opposing movements within local regions. Existing predictors approximate such motion using locally rigid transformations, which suffices for joints but fails in deformation-rich areas. This motivates the need for adaptive strategies that segment the mesh and apply transformation models aligned with regional motion characteristics.

\begin{figure}[h]
    \centering
    \includegraphics[width=0.63\linewidth]{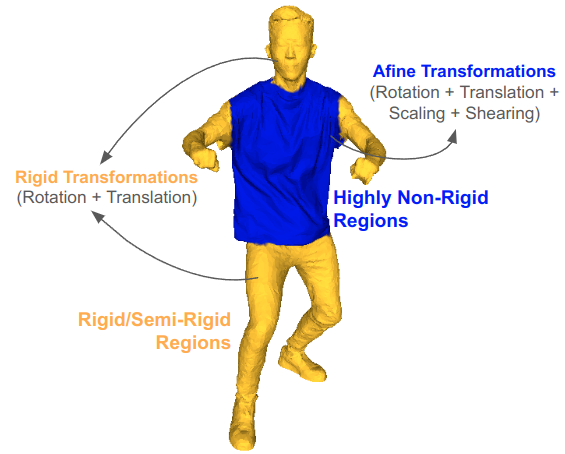}
    \caption{An illustration of the proposed Bi-modal Prediction Method. The method selectively applies a rigid transformation model to rigid/semi-rigid regions (e.g., limbs) and a more expressive affine transformation model to complex regions (e.g., loose clothing), guided by semantic segmentation.}
    \label{fig:bimodal_illustration}
\end{figure}

This paper advances KeyNode-driven compression through prediction flexibility for complex and non-rigid motion. This work proposes two main contributions. First, to overcome the performance limitations observed in existing KeyNode-driven approaches when faced with complex deformations, {\it Enhanced Prediction Flexibility via a Bi-modal Method} is presented. This approach leverages segmentation information to adaptively apply different types of transformations across the mesh (Fig. \ref{fig:bimodal_illustration}). It extends conventional rigid transformations (rotation \& translation) to incorporate more versatile affine transformations (including scaling and shearing) in regions identified as exhibiting complex, highly non-rigid motion. For optimized bitrate control, these affine matrices are decomposed into their fundamental components (rotation, translation, scaling, and shearing). A rate-distortion guided selection strategy, employing a Lagrangian formulation, is used to identify the optimal combination of these decomposed affine components, balancing between reconstruction quality and data rate.

Second, recognizing the spatial and temporal correlations present in translation vectors in dynamic mesh sequences, {\it Translation Predictive Coding Techniques} are proposed to reduce motion vector redundancy. This is achieved with Spatial Prediction and Spatio-Temporal Prediction. Spatial Prediction, designed for initial P-frames or when temporal information is less stable, uses a top-down spiral traversal order on a spatially constructed graph to predict translations based on neighboring nodes. Spatio-Temporal Prediction, exploits frame-to-frame correlations, predicting a node's translation based on the decoded temporal change of its corresponding body part in the previous frame. 

These proposed methods collectively advance the compression performance and adaptability of KeyNode-driven codecs, facilitating more efficient and accurate representation of diverse real-world scanned 3D human dynamic meshes, especially in bandwidth-constrained environments and for highly intricate motions.  The contributions of this work are as follows.
\begin{itemize}
    \item We extend the conventional rigid prediction framework with a hybrid scheme that selectively applies affine transformations to deformation-rich regions exhibiting pronounced non-rigid variability. This semantic-aware approach reduces the number of transmitted parameters by limiting extra modeling flexibility to the areas that need it, while preventing unnecessary complexity in simpler regions.
    \item For complex regions, affine transformations are factorized into elementary components, and only the essential components are encoded, thereby reducing bitrate overhead.
    \item A Lagrangian rate–distortion optimization strategy guides the component selection process, balancing compression efficiency and reconstruction quality.
    \item We propose two predictive coding approaches for translation vectors—Spatial and Spatio-Temporal Predictive Coding—which exploit inherent spatial and temporal correlations to further improve coding efficiency.
\end{itemize}

\section{Related Work}
\label{sec:related_work}

This section reviews existing literature relevant to dynamic mesh compression, highlighting the complexities of real-world scanned data.

\subsection{Dynamic Mesh Compression: Challenges and Paradigms}

The compression of dynamic 3D meshes faces multifaceted challenges due to the immense data volumes \cite{4373329}. A primary difficulty stems from the varying topology of real-world scanned meshes, which often exhibit inconsistent vertex counts, connectivity, and correspondence across frames, hindering the exploitation of temporal redundancies \cite{Survey_TVM_2025, Hoang_2025_KeyNode}. Furthermore, these meshes are susceptible to scan artifacts like holes and noise, complicating prediction and compression \cite{Hoang_2025_KeyNode}. Human motion itself presents a complex interplay of rigid and non-rigid deformations, particularly in areas like loose clothing, which are harder to model and compress than purely rigid movement. This complexity, coupled with the intertwined nature of shape and sampling components in 3D coordinates, necessitates the development of more adaptive and robust compression frameworks that can handle these complexities without extensive preprocessing, potentially through a holistic co-design of reconstruction and compression pipelines.

\begin{figure}[h]
    \centering
    \includegraphics[width=0.95\linewidth]{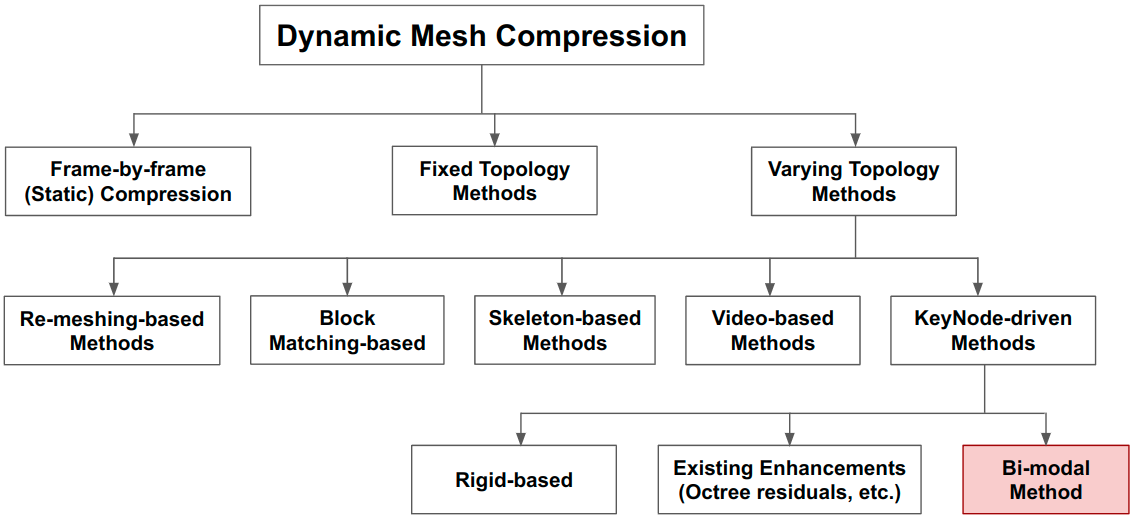}
    \caption{Taxonomy of dynamic mesh compression methods, including static, fixed topology, and varying topology approaches.}
    \label{fig:taxonomy}
\end{figure}

3D dynamic mesh compression methods fall into three categories: frame-by-frame (static), fixed topology, and varying topology (Fig.~\ref{fig:taxonomy}). Frame-by-frame approaches compress each mesh independently using schemes such as corner table \cite{923399} or TFan \cite{Mamou2009}, but they fail to exploit temporal coherence and thus have limited compression efficiency. Fixed topology methods assume vertex correspondences remain consistent across frames, enabling predictive coding or Principal Component Analysis (PCA)-based techniques to exploit inter-frame correlations~\cite{10.5555/789086.789628, 10.1006/cviu.2002.0987, SPC, BICI2011577, 6387603, 10.1111:1467-8659.00433, Karni2004, 8486541, 10.1145/1073368.1073398, 10.1145/1028523.1028547, Payan2007, Fast_Spatio_Temporal_2021}. 

Varying topology methods address real-world scanned meshes where connectivity changes and vertex correspondences are unavailable. Re-meshing approaches enforce a consistent global topology~\cite{1464745, 10.1145/2766945}; block-matching methods extend 2D motion estimation into 3D~\cite{4358669, 5652911, 1239447}; skeleton-based methods use skeleton tracking to remove temporal redundancy \cite{6804649, 6854785}; and video-based schemes map meshes to images for standard video encoding. The latter dominates with the MPEG V-DMC standard~\cite{9922888}, which represents base and subdivided meshes via displacement fields and video compression, further improved by refinements that enhance temporal and spatial compression performance~\cite{10222117, 10447762, 10402693, 10566382, 10772848, 10772781, 10647339, 10647545, 10647600, 10648035}. KeyNode-driven methods model motion through weighted transformations of sparse control nodes~\cite{10350716, Hoang_2025_KeyNode}. Extensions introduced octree-based residuals \cite{Hoang_2025_KeyNode}, bidirectional prediction \cite{Hoang_2025_KeyNode}, and compression-aware node placement~\cite{10350716}. Our Bi-modal method belongs to this class, designed to mitigate the rigidity of prior models by enabling more flexible prediction.

\subsection{KeyNode-Driven Compression and its Evolution}

\subsubsection{Foundational Principles}

The core principle of KeyNode-driven methods involves formulating the temporal motion of each vertex as a distance-weighted combination of transformations derived from a sparse set of control nodes \cite{Hoang_2025_KeyNode, 10350716}. This strategy achieves substantial data reduction by requiring the transmission of solely the transformations associated with these key nodes, rather than the full vertex positions for every frame \cite{Hoang_2025_KeyNode}.

A common underlying mechanism for computing these KeyNode-driven motion vectors is \textbf{Embedded Deformation (ED)}. Various ED methods have been proposed to capture dynamic motion \cite{sumner2007embedded, li2009robust, Chen2022}. Some are designed for complex human motions, such as a hybrid model that combines embedded and isometric deformation to capture both rigid and non-rigid complexities in human motion \cite{Chen2022}. Using ED, the geometry of a subsequent frame can be predicted by deforming the previously decoded frame based on the transformations of the embedded key nodes \cite{Hoang_2025_KeyNode}.

\subsubsection{Strengths and Limitations}

KeyNode-driven methods are highly effective for predicting rigid body motion and for modeling mesh regions that undergo limited non-rigid deformations, such as joints. A significant advantage of this deformation-based approach is its robustness to inconsistent topology and unreliable vertex correspondence, common issues in real-world scanned data \cite{Hoang_2025_KeyNode}. By focusing on the motion of a sparse set of controlling key nodes, the method can effectively capture the mesh's dynamics even when its topology varies across frames, overcoming the limitations of frame-to-frame differencing \cite{Hoang_2025_KeyNode}. This approach offers more flexibility than block-based methods--which assign a single motion vector to an entire spatial region--as each vertex can move independently without being bound to rigid spatial groupings. KeyNode-driven approaches allow for fine-grained, non-rigid deformation with per-vertex flexibility, while still maintaining a compact representation \cite{Hoang_2025_KeyNode}.

However, reliance on rigid transformations has limited effectiveness in strongly non-rigid regions such as loose clothing.
In principle, such motion could be approximated by dramatically increasing the density of the node graph, but this comes at the cost of higher bitrate and reduced compression efficiency. 
The existing enhancements, such as octree-based residual coding or dual-direction prediction, primarily serve to refine the rigid prediction or manage error accumulation within a rigid framework, without addressing its inefficiency for strongly non-rigid regions. This motivates the need for adaptive strategies that partition the mesh into regions and apply transformation models aligned with their motion characteristics.

Another challenge in predictive coding is the accumulation of distortion across frames \cite{10350716}. Later predicted frames (P-frames) typically exhibit higher distortion compared to earlier ones due to the propagation of prediction errors \cite{10350716}.  This accumulation becomes even more severe when the prediction model itself lacks sufficient flexibility for motion prediction. In particular, while the original KeyNode-driven methods offer generality, the reliance on a universally applied, relatively simple deformation model (rigid transformations) limits their effectiveness. Such rigid models cannot adequately capture highly complex, non-rigid deformations, such as those seen in loose clothing, thereby compounding the distortion problem.

\subsection{Advanced Deformation Models for Non-Rigid Motion}
A wide range of advanced deformation models has been developed in computer graphics and vision, such as free-form deformation \cite{10.1145/237170.237247} and cage-based animation \cite{Strter2024}. These approaches excel at capturing highly non-rigid behaviors and fine-grained deformations beyond the reach of rigid or piecewise rigid motion models. However, their expressive power comes with substantial computational and coding costs: parameter counts are high, models often require dense correspondences or volumetric grids, and incorporating them into compression pipelines can drastically increase bitrate and inference overhead.

\subsubsection{Affine Transformations in 3D Data Processing}

Affine transformations are extensively used for manipulating objects in 3D space due to their ability to preserve parallelism of lines and ratios of distances, while allowing for changes in shape, size, and orientation \cite{hartley2003multiple, foley2014computer}. 
Multi-dimensional affine transformations have also been applied in image and video compression for motion compensation, improving accuracy and efficiency over purely translational models \cite{USPatent_5970173A_1999}.

In the context of mesh deformation, affine transformations are foundational. Some methods propose mesh deformation based on affine transformation, often in conjunction with skeleton transformations, with the aim of reducing the need for remeshing in applications like electromagnetic inverse problems \cite{Mesh_Deformation_Affine_2024}. Piecewise affine prediction can also be employed for dynamic 3D meshes, where the motions of vertex clusters are described by a single affine motion model \cite{Piecewise_Affine_2008}.

While affine transformations are well-established for general 3D manipulation and motion compensation, their application to 3D dynamic mesh compression is a less explored avenue. To our knowledge, existing mesh deformation methods using affine transformations focus on modeling flexibility or remeshing reduction rather than the bitrate-optimized coding of the affine parameters themselves for dynamic mesh sequences. This highlights an opportunity: leveraging the expressive power of affine transformations not universally, but strategically. By selectively applying affine models only to motion-intensive, non-rigid regions, and efficiently encoding their decomposed parameters, compression efficiency gains can be achieved without over-modeling simpler areas.

\subsubsection{Neural Representations}
Neural representations have been proposed for capturing complex non-rigid 3D motion, including Neural Deformation Graphs and implicit neural shape models \cite{Neural_Deformation_Graphs_2020, Merrouche_2024_NeuralFields}.
While these neural methods excel at \textit{reconstructing} complex non-rigid motion and generating novel views, their integration into \textit{compression} pipelines for explicit mesh data remains challenging \cite{Chen_2024_GaussianSplatting}, due to high computational cost, difficulty in bitrate control, and the need for dense correspondences.

The proposed method, by extending KeyNode-driven frameworks with segmentation-aware affine transformations, exemplifies a hybrid approach. It enhances the explicit mesh compression framework with more sophisticated deformation capabilities without fully committing to the implicit neural paradigm, which tends to introduce new challenges regarding interpretability, computational overhead for explicit mesh extraction, and direct bitrate control. This highlights a pragmatic and strategic pathway for advancing explicit mesh compression by integrating advanced concepts where they offer the most targeted benefits, thus providing a practical solution for dynamic mesh compression.

\subsection{Summary}

Dynamic 3D human mesh compression is challenged by the irregularities of real-world scans and the complexity of non-rigid motion. We introduce a Bi-modal Prediction Method that adaptively combines simple rigid models with expressive affine transformations guided by semantic segmentation, enabling compression to better align with local motion characteristics. This hybrid design offers a scalable path toward practical, high-fidelity compression of dynamic meshes.

\section{Proposed Bi-modal KeyNode-Driven Codec}
\label{sec:codec_architecture}

\begin{figure}[h]
\centering
\includegraphics[width=\linewidth]{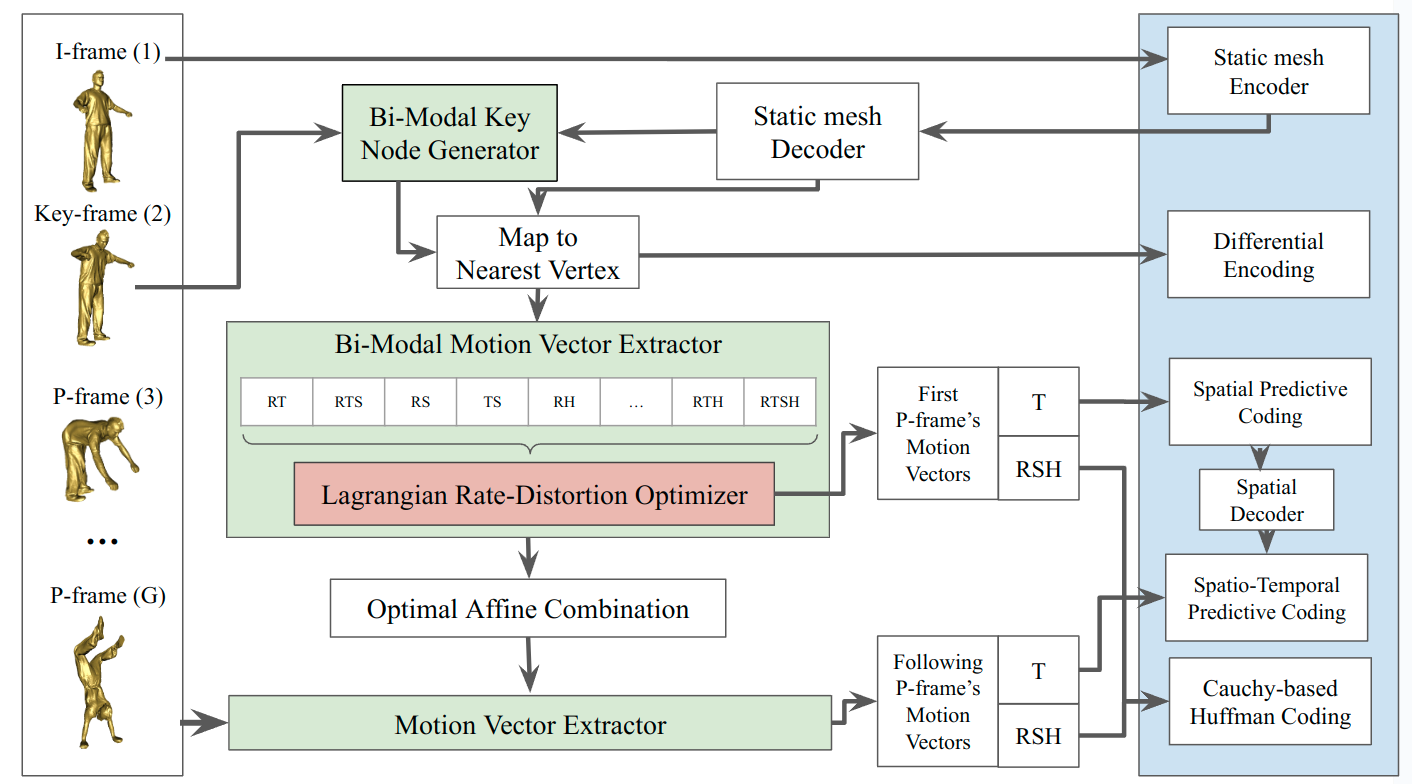}
\caption{The Bi-modal Encoder pipeline. The encoder leverages an enhanced KeyNode generator, decomposes affine components into rotation (R), translation (T), scaling (S), and shear (H) -- creating multiple combinations, and a rate-distortion optimizer to efficiently choose motion vectors for P-frames.}
\label{fig:encoder}
\end{figure}

\begin{figure}[h]
\centering
\includegraphics[width=\linewidth]{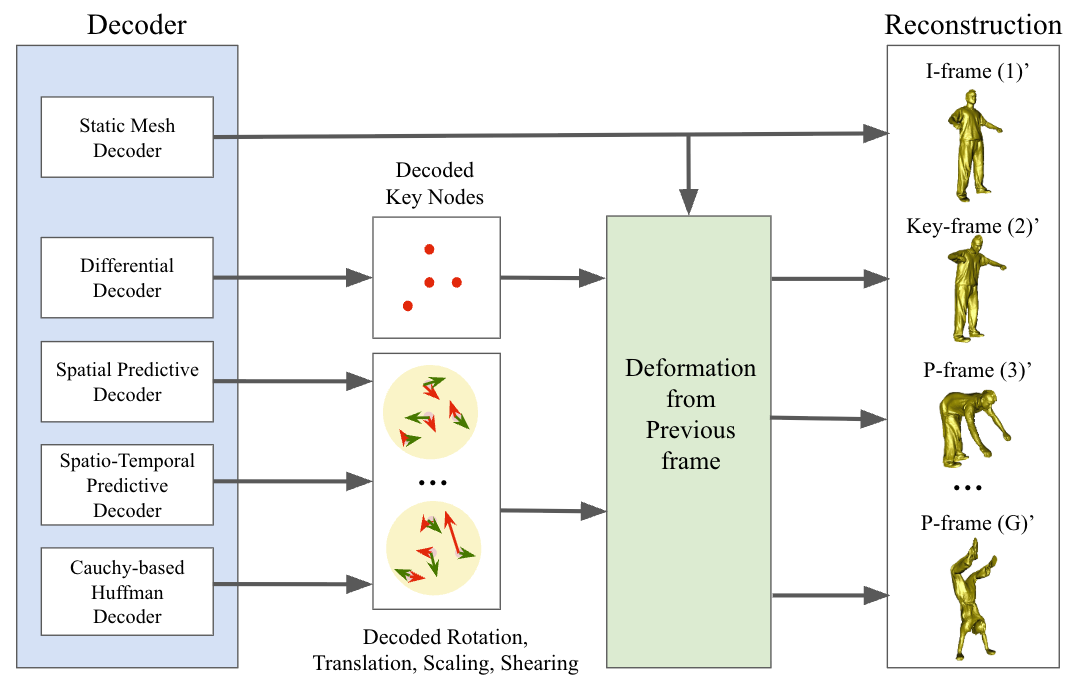}
\caption{The Bi-modal Decoder pipeline. The decoder reconstructs I-frames and key nodes, then uses predictive decoders and the Bi-modal deformation model to reconstruct subsequent P-frames.}
\label{fig:decoder}
\end{figure}

To enable efficient transmission, we have designed a complete system that integrates predictive motion modeling with specialized coding strategies. The proposed encoder and decoder structures, illustrated in Fig. \ref{fig:encoder} and Fig. \ref{fig:decoder}, respectively, form the backbone of our codec.

\subsection{System Architecture}
\label{sec:system_arch}

The encoding process begins by partitioning the input dynamic mesh sequence into Groups of Frames (GoFs), each starting with an I-frame. A key P-frame is selected within each GoF to serve as the reference for extracting key nodes. This task is handled by our \textbf{Bi-modal KeyNode Generator}, an enhanced version of the original design that accommodates not only rigid motion but also affine transformations for highly non-rigid regions (Section~\ref{sec:BiModalKeyNodeGenerator}).

Unlike the original KeyNode-driven codec, which relies on a fixed motion model, our Bi-modal codec introduces a \textbf{Bi-Modal Motion Vector Extractor} (Section~\ref{sec:affine_decomposition}) that evaluates multiple transformation types and identifies the most suitable representation for each GoF. The final choice is made by a \textbf{Lagrangian Rate-Distortion Optimizer} (Section~\ref{sec:Lagrangian}), ensuring the optimal trade-off between compression efficiency and geometric fidelity. The selected transformation combination is consistently applied to all P-frames within the GoF. For the first P-frame of a GoF, translation vectors (T) are encoded using \textbf{Spatial Predictive Coding} (Section~\ref{sec:spatial_pred}); for subsequent P-frames, \textbf{Spatio-Temporal Predictive Coding} (Section~\ref{sec:spatio-temporal-pred}) exploits temporal correlations. The remaining affine components—rotation (R), scaling (S), and shearing (H) (Section~\ref{sec:affine_decomposition})—are compactly represented using \textbf{Cauchy-based Huffman Coding} (Section~\ref{sec:mv_coding}).

The decoder mirrors this architecture. The I-frame geometry is first decoded using a static mesh decoder, while the key nodes are reconstructed with a differential decoder. Subsequently, for each P-frame, the affine transformation parameters are recovered from the bitstream and applied to the key nodes. The decoded parameters for rotation, scaling, and shearing are recovered using a Cauchy-based Huffman decoder, while the translation vectors are reconstructed using either the Spatial or Spatio-Temporal Predictive Decoders, a choice dictated by the coding mode. These decoded parameters are then used to reconstruct the final P-frame geometry through the Bi-Modal deformation model, using the previously decoded frame as a reference.

\subsection{Bi-Modal Key Node Generator}
\label{sec:BiModalKeyNodeGenerator}

To enhance stability and efficiency, our Bi-modal KeyNode Generator introduces two key improvements over the baseline method \cite{Hoang_2025_KeyNode}.

\subsubsection{Integrated Node Refinement}
In the baseline KeyNode Generator method \cite{Hoang_2025_KeyNode}, the deformation process was treated as a black-box optimization invoked iteratively within each node pruning step. This nested-loop structure led to significant computational overhead. Our new approach directly integrates node refinement into the primary deformation loop. This architectural change eliminates the redundant nested optimization, resulting in a more efficient and streamlined generation process. The node generation process is run for an increased number of iterations to ensure full convergence, with the initial key node set sufficiently dense to maintain pruning accuracy post-refinement.

\subsubsection{Improved Node Removal Strategy}
The original node removal strategy allowed for the simultaneous pruning of multiple low-error nodes, which could lead to instability if those nodes were in the same region of influence. To address this, we introduce a filtering mechanism that prevents the simultaneous removal of nodes that affect the same vertex (defined by a control weight greater than 0.01). This new strategy significantly reduces interference during pruning and improves overall stability. Additionally, we incorporate a node insertion capability to further enhance solution quality. If the current number of nodes is less than a target count, new nodes are inserted at vertex locations exhibiting the highest MSDM2-error, ensuring optimal distribution of key nodes. A distance threshold is used to prevent newly added nodes from being placed too close to existing ones.

\subsection{Rate-Distortion Optimization}\label{sec:Lagrangian}

To select the best affine combination for a given motion, we use a Rate-Distortion optimizer based on a Lagrangian Formulation. With multiple possible combinations of affine components, each combination will have its own final rate and distortion. The Lagrangian cost function is:

\begin{equation}
J(\Omega, \lambda) = D(\Omega) + \lambda R(\Omega)
\end{equation}
where $D(\Omega)$ is the distortion and $R(\Omega)$ is the number of bytes associated with transmitting the affine transformation $\Omega$. The optimal affine transformation, $\widehat{\Omega}$, is the one that minimizes this cost function:

\begin{equation}
\widehat{\Omega} = \underset{\Omega \in \Phi}{\operatorname{argmin}} J(\Omega, \lambda) = \underset{\Omega \in \Phi}{\operatorname{argmin}} (D(\Omega) + \lambda R(\Omega))
\end{equation}
where $\Phi$ is the set of all possible affine combinations.

\section{The Bi-modal Prediction Framework}
\label{sec:bi-modal-pred}

While the codec establishes the overall system pipeline, the \textbf{Bi-modal Prediction Framework} provides the core mechanisms that make compression effective.

\subsection{The Representation Gap of Rigid Models}

The original KeyNode-Driven Codec shows significant compression performance, especially for slow-motion sequences. However, its performance degrades as the motion becomes more complex. The prediction method \cite{Chen2022} addresses some semi-rigid transformations in human bodies, such as joint regions, by dividing the mesh into clusters and using key nodes to simulate non-rigid deformation through locally rigid transformations. Simulating non-rigid transformations as locally rigid is a common approach \cite{6361384} because rigid registration has fewer degrees of freedom (DoF), making it computationally feasible to solve. While this approach works well for meshes with limited non-rigid deformations, such as those in joints, it performs poorly in loose clothing regions.

\subsection{Bi-modal Affine Transformations for Non-Rigid Regions}

To address the representation gap for non-rigid motion, we propose to use semantic \textbf{segmentation} information and extend the types of transformations applied to key nodes, increasing the model's ability to represent motion more flexibly.

The deformation model is a central part of our framework, which uses affine transformations for non-rigid regions and rigid transformations elsewhere. The deformation is expressed as a weighted sum of transformations from neighboring key nodes.
Let $A_n \in \mathbb{R}^{3 \times 3}$ be the affine matrix and $t_n \in \mathbb{R}^3$ be the translation vector associated with a key node $n$. The deformation is then expressed as:

\begin{equation}
x_i' = \sum_{j=1}^{Q} w_j \left[ A_j(x_i - n_j) + t_j + n_j \right]
\end{equation}
\noindent where $x_i$ is the original vertex position, $x_i'$ is the deformed vertex position, $n_j$, $A_j$, $t_j$ are the neighboring key node and its affine matrix and translation vector, respectively, and $Q$ is the number of nodes controlling the deformation of a vertex. The weights $w_j$ are defined as

\begin{equation}
    w_j = \frac{1}{d^2(x_i, n_j)}
\end{equation}

\noindent and are subsequently normalized such that $\sum_j w_j = 1$, 
ensuring that closer nodes exert proportionally greater influence.

From a compression standpoint, it is inefficient to allocate extra bits to areas where motion remains rigid. We apply the affine transformation only in regions with complex clothing movement (specifically, the torso/T-shirt area) while keeping rigid transformations elsewhere. This prediction method is therefore called \textit{Bi-modal}. 

The affine matrices are not inherently constrained to be orthogonal. To ensure the transformation preserves a natural structure while allowing for flexibility, an additional orthogonality loss is introduced to encourage the affine matrix to remain close to an orthogonal matrix:
\begin{align}
\mathcal{L}_{orth} = \sum_j \Big(
    (a_1^T a_2)^2 + (a_1^T a_3)^2 + (a_2^T a_3)^2 \nonumber \\
    + (1 - a_1^T a_1)^2 + (1 - a_2^T a_2)^2 + (1 - a_3^T a_3)^2
\Big)
\end{align}
\noindent where $a_1, a_2, a_3$ are the column vectors of $A_j$.
This helps prevent the affine matrix from introducing excessive scaling or shearing in regions where it is not needed.
The optimal values for $A$ and $t$ are obtained through an optimization process, formulated as:

\begin{equation}
\underset{A, t}{\mathrm{argmin}} (\mathcal{L}_{data} + \alpha_{reg} \mathcal{L}_{reg} + \alpha_{orth} \mathcal{L}_{orth})
\end{equation}
\noindent Here, $\mathcal{L}_{data}$ penalizes the geometry deviation between source and target, and $\mathcal{L}_{reg}$ ensures the smoothness of the deformation.

In the Bi-modal method, segmentation labels are not required to be perfect. We provide two modes: \textit{auto-segmentation} (output of any conventional 3D segmentation method) and \textit{manual-segmentation} (human corrector). In\textit{ manual-segmentation} mode, we assume that segmentation is perfect, so only neighboring nodes with the same segmentation labels are allowed to control the deformation of a vertex. If fewer than $Q$ valid neighbors are available for a vertex $v_i$, the missing node(s) is supplemented through propagation from the existing valid nodes. When there are too few valid nodes in the desired body part(s), the deformation of a vertex is determined solely by the available valid nodes. For instance, if only one valid node is present, it is used $Q$ times; if two are available, the closest node is replicated $Q-1$ times, etc. Consequently, this method only fails when no valid nodes exist in the specified body part(s).
In \textit{auto-segmentation} mode, we allow the deformation process to continue even when no valid nodes are found after segmentation-based filtering. In such cases, the system falls back to the default behavior: it deforms the vertex using the $Q$ nearest nodes based on Euclidean distance, ignoring segmentation information.

\underline{\textit{Segmentation-enhanced Correspondence Estimation}}: We leverage segmentation information to establish a rough alignment between corresponding body parts prior to deformation prediction. This is crucial as original deformation prediction approaches, which rely on the Iterative Closest Point (ICP) method at their core, can struggle with large or rapid deformations. By estimating initial rigid transformations at the segment level using bounding volumes, we enhance the robustness of correspondence estimation, especially in scenarios involving fast motion or significant positional changes. However, as will be seen in the evaluation later on, this feature is not always helpful, especially in cases where the segmentation is inaccurate and/or inconsistent between frames.

\subsection{Affine Decomposition}
\label{sec:affine_decomposition}

To optimize bitrate, we avoid transmitting the full 9-parameter affine matrix. Instead, we decompose the affine matrix $A \in \mathbb{R}^{3 \times 3}$ into Rotation ($R \in \mathbb{R}^{3 \times 3}$), Scaling ($S \in \mathbb{R}^{3 \times 3}$), and Shearing ($H \in \mathbb{R}^{3 \times 3}$) and selectively encode only what is necessary:
\begin{equation}
A = R \times S \times H
\end{equation}

This decomposition is crucial for bitrate optimization, as it allows for flexible, variable-rate encoding where we only transmit the components necessary to represent a given motion. For example, a motion that is primarily rotational and translational can be encoded with just the rotation component and the translation vector. The decision of which components to enable is made on a per-motion basis using Rate-Distortion Optimization (Section \ref{sec:Lagrangian}) to select the optimal combination to balance visual quality and compression bitrate.

The decomposed components are represented as follows, each requiring only a minimal set of values for transmission.

{\it Rotation} in 3D is commonly represented using Euler angles ($\theta_x$, $\theta_y$, $\theta_z$), which define rotations around the X, Y, and Z axes. The rotation matrix is constructed as:
\begin{equation}
R = R_z(\theta_z) R_y(\theta_y) R_x(\theta_x)
\end{equation}
where:
\begin{equation}
R_x(\theta_x) =
\begin{bmatrix}
1 & 0 & 0 \\
0 & \cos{\theta_x} & -\sin{\theta_x} \\
0 & \sin{\theta_x} & \cos{\theta_x}
\end{bmatrix}
\end{equation}

\begin{equation}
R_y(\theta_y) =
\begin{bmatrix}
\cos{\theta_y} & 0 & \sin{\theta_y} \\
0 & 1 & 0 \\
-\sin{\theta_y} & 0 & \cos{\theta_y}
\end{bmatrix}
\end{equation}

\begin{equation}
R_z(\theta_z) =
\begin{bmatrix}
\cos{\theta_z} & -\sin{\theta_z} & 0 \\
\sin{\theta_z} & \cos{\theta_z} & 0 \\
0 & 0 & 1
\end{bmatrix}
\end{equation}

{\it Scaling} affects the size of an object along each axis. The scaling matrix $S$ is a diagonal matrix:

\begin{equation}
S =
\begin{bmatrix}
s_x & 0 & 0 \\
0 & s_y & 0 \\
0 & 0 & s_z
\end{bmatrix}
\end{equation}

{\it Shearing} skews an object in one direction while keeping the other coordinates fixed. It is represented as an upper triangular matrix $H$:

\begin{equation}
H =
\begin{bmatrix}
1 & h_{xy} & h_{xz} \\
0 & 1 & h_{yz} \\
0 & 0 & 1
\end{bmatrix}
\end{equation}

A total of 16 possible combinations can be formed from these four decomposed components (Rotation, Translation, Scaling, and Shearing). In our experiments, we will refer to the combination code as a flag that indicates which components are enabled, and the combination's name will be derived from the matrix letter of the enabled components. For example, a combination of Rotation, Scaling, and Shearing would have the code "1011" and the name "RSH."

\begin{itemize}
    \item \textbf{When Rotation is Disabled:} The rotation matrix is an identity matrix ($R=I$), by setting the Euler angles $\theta_x= \theta_y= \theta_z=0$.
    \item \textbf{When Translation is Disabled:} The translation vector is zero ($T = [0, 0, 0]$).
    \item \textbf{When Scaling is Disabled:} The scaling matrix $S=I$, by setting $s_x= s_y= s_z=1$.
    \item \textbf{When Shearing is Disabled:} The shear matrix $H=I$, by setting the shear parameters $h_{xy} = h_{xz} = h_{yz} =0$.
\end{itemize}

To reduce complexity, we simplify our Bi-Modal Motion Vector Extractor by always enabling Translation (T). This is a deliberate design choice because translation is a fundamental component of all motion. This reduces the number of affine combinations to consider from 16 to 8.

\subsection{Coding of Transformation Components}
\label{sec:mv_coding}
Let $N$ be the total number of key nodes. Let $N_a$, $N_r$ be the number of \textit{affine}-nodes and \textit{rigid}-nodes, respectively, with $N_a + N_r = N$. The type of each node is determined by its local context: for each node, the closest mesh vertex is identified, and the node inherits the segmentation label of that vertex.

Since we keep the model of using R and T for \textit{rigid}-nodes, there would be $6N_r$ values for \textit{rigid}-nodes. For \textit{affine}-nodes, the number of values to be transmitted will vary depending on which components are enabled.

For scaling S, we subtract 1 from $S_x$, $S_y$, $S_z$ and encode the residuals, denoted by $S' = [S_x - 1, S_y - 1, S_z - 1]$. Let $R' = [\theta'_x, \theta'_y, \theta'_z]$ and $H' = [h'_{xy}, h'_{xz}, h'_{yz}]$. Since we observe that the distributions of R', S', and H' are similar, we concatenate them into a single vector $P = [R', S', H']$. If any component is disabled, we simply omit it from P.

We encode P using Cauchy-based Huffman coding as proposed in \cite{Hoang_2025_KeyNode}, while T is encoded using various Predictive Coding schemes, as proposed below.

\subsection{Translation Predictive Coding}
\label{sec:translation_pred_coding}
Based on the noticeable and consistent patterns we've identified in Translation vectors across both space and time, we're using these trends and relationships to improve how we encode Translations through prediction. The original coding approach \cite{Hoang_2025_KeyNode} simply quantizes the original Translation vectors and then applies Huffman coding. To enhance this, we introduce new prediction techniques that capitalize on the spatial and spatio-temporal correlations that were previously observed: Spatial Prediction and Spatio-Temporal Prediction.
\subsubsection{Spatial Prediction}\label{sec:spatial_pred}
This method is designed for the initial P-frames, where no temporal information is available, or when the temporal information is less stable or inconsistent comparing to the spatial relationships. Spatial  Prediction mainly depends on the traversal order, which should ensure that consecutively traversed nodes have similar values. Another essential aspect of Spatial Prediction is the prediction model, which, along with the traversal order, will be described further below.

Given that nearby nodes often have similar Translation values, we aim to explore and make use of this trend. The key is to find an effective traversal order that captures this spatial relationship, ensuring that nodes close to each other are processed in a way that maximizes the prediction accuracy. 

\begin{figure}[h]
\centering
\includegraphics[width=0.3\textwidth]{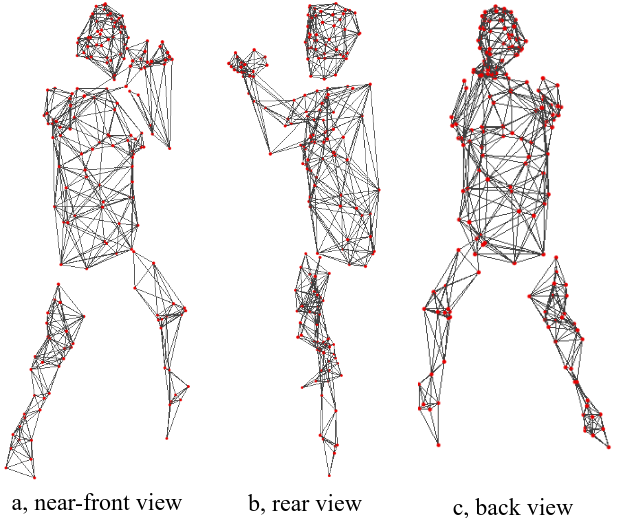}
\caption{The influence graph of key nodes.}
\label{fig:node_graph}
\end{figure}

\paragraph{Spatial Graph Construction}
Proximity is determined based on the shared influence on vertices. Specifically, if a vertex $v$ is deformed along with nodes $n_i, n_j, n_k$, these three nodes are considered connected. These connections will later be utilized for node traversal. Fig. \ref{fig:node_graph} provides a visualization of the constructed graph, where the red dots represent key nodes, and the black lines indicate the connecting edges. 

Since we are using the Bi-modal method, where the torso is deformed independently, the constructed graph in Fig. \ref{fig:node_graph} consists of isolated components: the head, torso, two arms, and two legs.

\begin{figure}[h]
\centering
\includegraphics[width=0.25\textwidth]{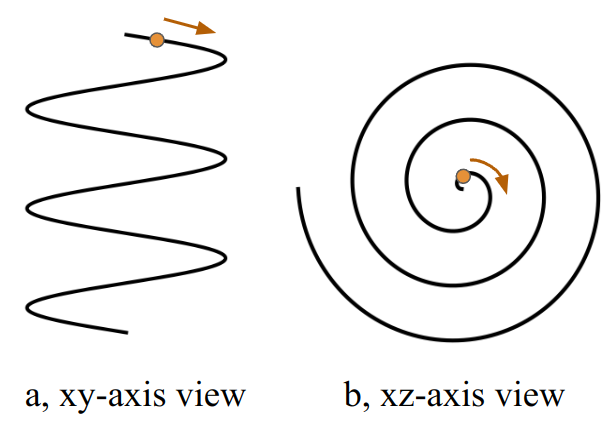}
\caption{Illustration of the spiral traversal order}
\label{fig:spiral_illustration}
\end{figure}

\paragraph{Traversal Order: Top-down Spiral Traversal}
Each component is traversed using a "spiral" pattern to ensure all nodes are visited while preserving similar translations. We scan the component from top to bottom along the y-axis. At each y-level, we define a horizontal layer and traverse it in a spiral pattern along the x and z axes, as shown in Fig. \ref{fig:spiral_illustration}.

To define a horizontal layer, we use the 1-ring neighborhood of the previous layer, including all nodes directly connected to those in the previous layer. This is where the graph constructed in Fig. \ref{fig:node_graph} becomes useful. After identifying the nodes in the 1-ring layer, we arrange them in a clockwise order. To achieve this, we first remove the y-values of the nodes (mapping them to the x-z plane) and then calculate the centroid of the 2D ring. Finally, we compute the angle of each node relative to the centroid, and sort them based on the angles. 

Since we have some isolated components (head, torso, 2 arms, and 2 legs), each component is traversed individually. However, the starting point for each component and the order in which they are processed follow a specific segmentation order. The component order is as follows: Head, Left upper arm, Left lower arm, Left hand, Right upper arm, Right lower arm, Right hand, Torso, Left thigh, Left leg, Left foot, Right thigh, Right leg, Right foot. 

The head's starting point is the node with the highest y-value, while the Left upper arm, Right upper arm, and Torso each begin at the node closest to the head (measured using a point-to-plane metric). The Left thigh and Right thigh each start at the node closest to the torso. We define the set of starting nodes as $N_s={\{n_s, n_p\}_j}$, where $n_s$ is the starting node for the connected component $j$, and $n_p$ is the closest node on the reference body part (head or torso). This method of selecting starting nodes ensures a smooth transition in the node transformations across body parts. For other body parts, such as the lower arms and hands, where the starting node is not explicitly defined, the traversal follows a spiral order based on their connection to adjacent components (e.g., the Left lower arm and Left hand are traversed together as a connected component with the Left upper arm).

\paragraph{Prediction}
Given the traversed nodes and their associated Translation vectors, denoted by $T_{spiral} = \{t_i \in \mathbb{R}^3, 0 \leq i \leq N-1, i \in S\}$, where $S$ is the traversal path, we apply differential prediction on consecutive translations for every node except the starting nodes in $N_s$. For each traversed node (except starting nodes $n_s$ in $N_s$), we compute the value to be transmitted as:
\begin{equation}
    t'_i = t_i - \hat{t}_{i-1} 
\end{equation}
where $\hat{t}_{i-1}$ is the decoded translation of the previous node in the traversal path.

For the starting nodes $n_s$ in $N_s$, $t'_{i}$ is calculated as:
\begin{equation}
    t'_{i} = t_{n_s} - \hat{t}_{n_p} 
\end{equation}
where $\hat{t}_{n_p}$ is the decoded translation of $n_p$.

To get the decoded value of the previous node, $\hat{t}_{i-1}$, we must first define the quantization rule and Huffman table. To do that, we first compute the pseudo-differences by taking the difference between consecutive nodes' translations without relying on the decoded values, (i.e., $t''_i = t_i - t_{i-1}$). Next, we apply Cauchy-based Huffman coding to the set of pseudo-differences, denoted by $T'' = \{t''_i\}$, which generates the necessary quantization rule and Huffman table. This allows us to reconstruct the decoded differential Translation for previous node, denoted by $\hat{t}'_{i-1}$, which is needed to reconstruct $\hat{t}_{i-1}$.

Once $t'_i$ is computed, it undergoes quantization and Huffman coding based on the rule derived for $T''$.

\subsubsection{Spatio-Temporal Prediction}\label{sec:spatio-temporal-pred}
We also observed strong frame-to-frame correlations in translations. It suggests that the temporal change of a node is generally predictable for the changes of other nodes within the same frame. We assume that nodes within the same body part typically exhibit similar translation vectors. 

With that assumption, given $\hat{T}_{f-1}$ and $T_f$ for each body part $b$, we first determine the representative temporal change $d_0^b$ such that the deviation from decoded $\hat{d}_0^b$ to translations of the belonging nodes are minimized:
\begin{equation}
    d_0^b = \operatorname{argmin} \left( || \hat{d}_0^b - t_i^f || \right) 
\end{equation}
To solve this optimization problem, we use the \textit{'L-BFGS-B'} optimization algorithm, which is available in the Scipy Python library. Instead of using 32-bit floating point values for transmitting $d_0^b$, we take advantage of the Cauchy-based Huffman dictionary from the previous P-frame (assuming the \textit{spatio-temporal} prediction mode is always used after at least one P-frame encoded by \textit{spatial} prediction). Therefore, $\hat{d}_0^b$ is achieved by encoding and decoding $d_0^b$ using the Cauchy-based Huffman dictionary of the previous P-frame.

The predicted translation for each node is calculated as:
\begin{equation}
    t_i^p = \hat{t}_i^{f-1} + \hat{d}_0^b 
\end{equation}
where $\hat{d}_0^b$ represents the decoded temporal change for the body part that node $i$ belongs to.

Next, we calculate the prediction errors for each node:
\begin{equation}
    e_i = t_i^f - t_i^p 
\end{equation}
Finally, we apply Cauchy-based Huffman coding to the prediction errors $\{e_i\}$, and send them to the decoder.

\section{Evaluation}

\subsection{Setup}
To ensure consistency and allow comparison with the latest proposed KeyNode-driven method \cite{Hoang_2025_KeyNode}, we adopt the same evaluation setup based on scanned dynamic mesh sequences. These sequences, summarized in Table~\ref{tab:dataset-details}, capture a wide range of motion patterns and reconstruction artifacts. In contrast to \cite{Hoang_2025_KeyNode}, where sequences were grouped into three categories (\textit{Slow Motion}, \textit{Fast Motion}, and \textit{Join/Split}), in this paper we merge the latter two into a single \textit{Complex Motion} category. \textit{Slow Motion} sequences exhibit high temporal coherence between consecutive frames, while \textit{Complex Motion} encompasses both rapid non-rigid body movements (e.g., dancing, waving clothing) and topological interactions (e.g., objects joining or separating), providing a challenge for compression methods.

\begin{table}[htb]
\caption{Dynamic mesh sequences evaluated. }
\resizebox{0.48\textwidth}{!}{
\begin{tabular}{|c|c|c|}
\hline
Category &
  Sequences &
  Characteristics \\ \hline
Slow Motion &
  Mitch, Soldier, Thomas &
  High temporal redundancy, smooth motion \\ \hline
Complex Motion &
  \begin{tabular}[c]{@{}c@{}}Longdress, Dancer, \\ Levi, Basketball\end{tabular} &
  \begin{tabular}[c]{@{}c@{}}Rapid non-rigid movements, \\ clothing dynamics, or object interactions\end{tabular} \\ \hline
\end{tabular}}
\label{tab:dataset-details}
\end{table}

For geometry distortion evaluation, we use the MSDM2 metric \cite{Lavou2011}, as in the earlier KeyNode-driven studies. MSDM2 measures curvature-based local geometric differences and has been validated through subjective experiments to correlate strongly with human perception. The metric outputs values in the range $[0,1]$, where lower values indicate higher similarity. Since MSDM2 is designed for static meshes, we report the average score across frames to assess sequence-level distortion.

\subsection{Rate-Distortion Performance Analysis}\label{sec:RD-comparison}
We evaluate the rate-distortion (RD) performance of our proposed Bi-modal  codec against two baselines: the earlier KeyNode-driven method \cite{Hoang_2025_KeyNode} (denoted \textit{TMM}), and the geometry coding method from V-DMC \cite{9922888}. For V-DMC, we include both \textit{all intra-coding} (denoted \textit{V-DMC--intra}) and  \textit{inter-frame coding} (denoted \textit{V-DMC--inter}).   

The Bi-modal method is evaluated in two segmentation modes: \textit{auto segmentation} (\textit{BiModal-A}) and \textit{manual segmentation} (\textit{BiModal-M}) when manual segmentation is available. This allows us to assess the influence of segmentation quality on coding efficiency. Auto-segmentation labels are generated using the 3D human segmentation approach proposed in \cite{Keito_3D}.
Manual labels are derived by editing the auto-segmented results, applying corrections when auto-segmentation violates body structure- for example, when body parts are misconnected (e.g., vertices of the right hand labeled as the left hand) or when multiple disconnected components share the same body part label. Minor boundary inaccuracies that do not conflict with the anatomical structure are left unaltered.

\subsubsection{Performance on Complex Motion Sequences}

\begin{figure}[H]
    \centering
    \includegraphics[width=\linewidth]{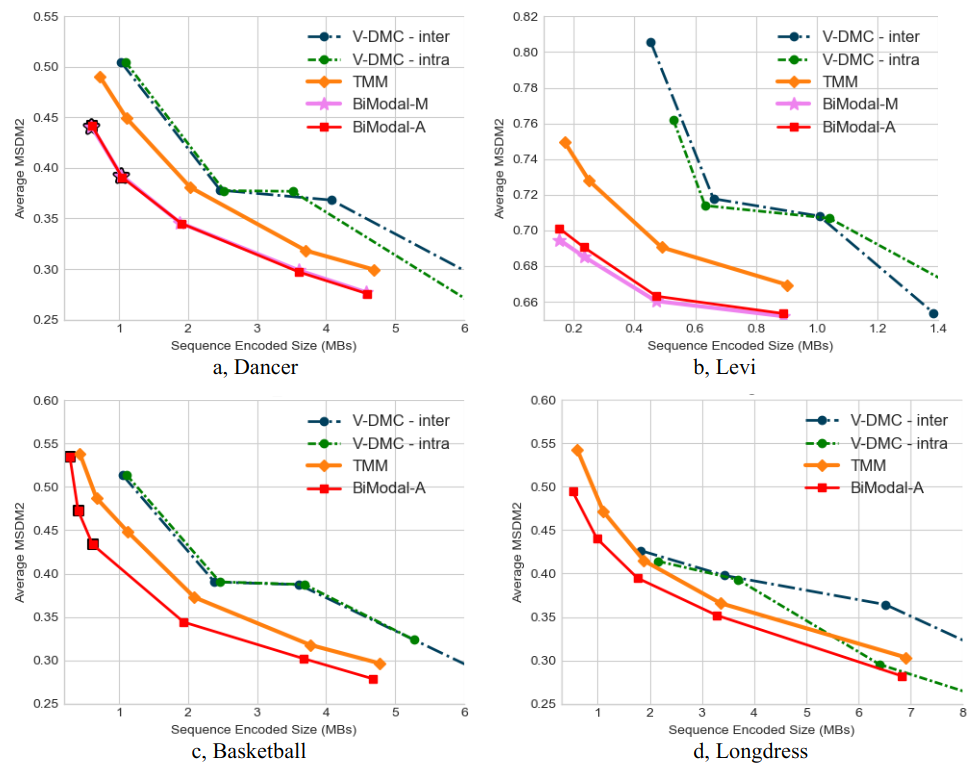}
    \caption{Rate–distortion performance on \textit{Complex Motion} sequences. Data points highlighted with black borders (appearing at the low end of some curves in parts (a) and (c)) indicate alternative configurations used to avoid GPU memory crashes.}
    \label{fig:rd-complex}
\end{figure}

Fig.~\ref{fig:rd-complex} shows the R–D curves for the \textit{Complex Motion} category, which includes Dancer, Levi, Basketball, and Longdress.

We aim to keep all configurations between TMM and Bi-modal consistent, 
except for the number of key nodes. The Bi-modal method requires fewer 
key nodes, since it can maintain acceptable quality with a reduced node 
count (more detail in Section \ref{sec:reduced_num_nodes}). However, a few data points on the far left of the Bi-modal curves for \textit{Dancer} and \textit{Basketball}—highlighted with black borders—do not correspond to identical configurations used by TMM. This discrepancy arises because the original settings caused GPU memory crashes. Specifically, these settings involved I-frames with a large number of vertices; while TMM only includes rotation and translation, the Bi-modal method also incorporates scaling and shearing, doubling the parameter space and exceeding GPU memory capacity. To address this, we used the closest viable configuration (represented by the middle dot) and increased the GoF size to achieve a lower bitrate. The same adjustment is applied consistently to both segmentation modes (\textit{BiModal-A} and \textit{BiModal-M}).  

The results indicate that Bi-modal achieves lower distortion at equivalent 
bitrates compared to TMM. For \textit{Dancer}, the two Bi-modal modes show nearly identical performance, since fewer than 5\% of frames require manual correction after auto-segmentation. By contrast, segmentation of \textit{Longdress} is challenging: the loose clothing obscures body parts, making reliable segmentation difficult, even with manual refinement. Consequently, the performance gain of \textit{BiModal-A} on \textit{Longdress} is smaller than for other complex-motion sequences, yet still achieves measurable improvements, suggesting Bi-modal's added flexibility remains beneficial even under poor segmentation. Overall, for sequences with highly non-rigid motion, Bi-modal provides a significant performance advantage over the baselines.

\subsubsection{Performance on Simple Motion Sequences}

\begin{figure}[h]
    \centering
    \includegraphics[width=\linewidth]{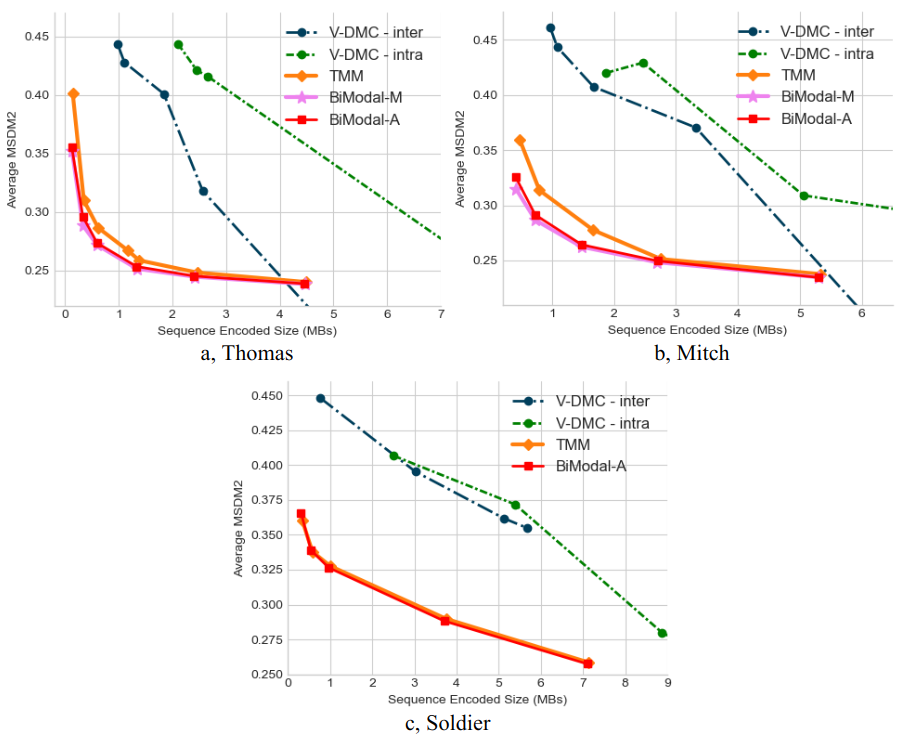}
    \caption{RD performance on \textit{Slow Motion} sequences.}
    \label{fig:rd-slow}
\end{figure}

Fig.~\ref{fig:rd-slow} presents the R–D curves for the \textit{Slow Motion} category. Here, the performance gap between Bi-modal and TMM is small. Since the improvements of Bi-modal are primarily designed to handle complex and fast non-rigid movement challenges, its advantage is less pronounced. In all cases where \textit{BiModal-M} is available, the difference between segmentation modes is minimal because only a small portion of frames require manual correction after auto-segmentation. Overall, the results indicate that Bi-modal provides only minor improvements for sequences with slow or simple motion, where temporal redundancy already favors prediction quality.

The R-D performance comparison can be summarized using 
BD-rate \cite{Bjntegaard2001CalculationOA}, which quantifies the bit rate difference needed to achieve the same quality. 
Table \ref{tab:BD-rate} presents the BD-rate percentages compared to the baselines; negative values indicate that Ours outperforms the comparison method, while positive values indicate that Ours performs worse.

\begin{table*}[h]
\caption{BD-rate (\%) of Ours against Baselines. \textcolor[HTML]{036400}{Green Values} represent BiModal-A, while \textcolor[HTML]{9a00d8}{Purple Values} represent BiModal-M.}
\centering
\renewcommand{\arraystretch}{1.5}
\resizebox{0.95\textwidth}{!}{
\begin{tabular}{c|clclclcl|clclcl|l}
\hline
 &
  \multicolumn{8}{c|}{\textbf{Complex Motion}} &
  \multicolumn{6}{c|}{\textbf{Slow Motion}} &
  \multicolumn{1}{c}{} \\ \cline{2-15}
\multirow{-2}{*}{} &
  \multicolumn{2}{c|}{Longdress} &
  \multicolumn{2}{c|}{Dancer} &
  \multicolumn{2}{c|}{Levi} &
  \multicolumn{2}{c|}{Basketball} &
  \multicolumn{2}{c|}{Soldier} &
  \multicolumn{2}{c|}{Thomas} &
  \multicolumn{2}{c|}{Mitch} &
  \multicolumn{1}{c}{\multirow{-2}{*}{\textbf{Avg.}}} \\ \hline
TMM &
  \multicolumn{1}{c|}{{\color[HTML]{036400} \textbf{-26.04}}} &
  \multicolumn{1}{l|}{{\color[HTML]{9a00d8} -}} &
  \multicolumn{1}{c|}{{\color[HTML]{036400} \textbf{-33.80}}} &
  \multicolumn{1}{l|}{{\color[HTML]{9a00d8} \textbf{-33.31}}} &
  \multicolumn{1}{c|}{{\color[HTML]{036400} \textbf{-50.75}}} &
  \multicolumn{1}{l|}{{\color[HTML]{9a00d8} \textbf{-56.31}}} &
  \multicolumn{1}{c|}{{\color[HTML]{036400} \textbf{-36.71}}} &
  {\color[HTML]{9a00d8} -} &
  \multicolumn{1}{c|}{{\color[HTML]{036400} \textbf{-4.4}}} &
  \multicolumn{1}{l|}{{\color[HTML]{9a00d8} -}} &
  \multicolumn{1}{c|}{{\color[HTML]{036400} \textbf{-25.81}}} &
  \multicolumn{1}{l|}{{\color[HTML]{9a00d8} \textbf{-34.94}}} &
  \multicolumn{1}{c|}{{\color[HTML]{036400} \textbf{-32.37}}} &
  {\color[HTML]{9a00d8} \textbf{-37.49}} &
  \textbf{-33.81} \\ \hline
V-DMC inter &
  \multicolumn{1}{c|}{{\color[HTML]{036400} \textbf{-44.39}}} &
  \multicolumn{1}{l|}{{\color[HTML]{9a00d8} -}} &
  \multicolumn{1}{c|}{{\color[HTML]{036400} \textbf{-41.89}}} &
  \multicolumn{1}{l|}{{\color[HTML]{9a00d8} \textbf{-41.58}}} &
  \multicolumn{1}{c|}{{\color[HTML]{036400} \textbf{-68.47}}} &
  \multicolumn{1}{l|}{{\color[HTML]{9a00d8} \textbf{-71.10}}} &
  \multicolumn{1}{c|}{{\color[HTML]{036400} \textbf{-47.06}}} &
  {\color[HTML]{9a00d8} -} &
  \multicolumn{1}{c|}{{\color[HTML]{036400} \textbf{-87.11}}} &
  \multicolumn{1}{l|}{{\color[HTML]{9a00d8} -}} &
  \multicolumn{1}{c|}{{\color[HTML]{036400} \textbf{-81.19}}} &
  \multicolumn{1}{l|}{{\color[HTML]{9a00d8} \textbf{-83.23}}} &
  \multicolumn{1}{c|}{{\color[HTML]{036400} \textbf{-76.76}}} &
  {\color[HTML]{9a00d8} \textbf{-76.32}} &
  \textbf{-65.37} \\ \hline
V-DMC intra &
  \multicolumn{1}{c|}{{\color[HTML]{036400} \textbf{-26.87}}} &
  \multicolumn{1}{l|}{{\color[HTML]{9a00d8} -}} &
  \multicolumn{1}{c|}{{\color[HTML]{036400} \textbf{-40.48}}} &
  \multicolumn{1}{l|}{{\color[HTML]{9a00d8} \textbf{-40.06}}} &
  \multicolumn{1}{c|}{{\color[HTML]{036400} \textbf{-71.48}}} &
  \multicolumn{1}{l|}{{\color[HTML]{9a00d8} \textbf{-74.94}}} &
  \multicolumn{1}{c|}{{\color[HTML]{036400} \textbf{-49.98}}} &
  {\color[HTML]{9a00d8} -} &
  \multicolumn{1}{c|}{{\color[HTML]{036400} \textbf{-77.66}}} &
  \multicolumn{1}{l|}{{\color[HTML]{9a00d8} -}} &
  \multicolumn{1}{c|}{{\color[HTML]{036400} \textbf{-92.70}}} &
  \multicolumn{1}{l|}{{\color[HTML]{9a00d8} \textbf{-93.64}}} &
  \multicolumn{1}{c|}{{\color[HTML]{036400} \textbf{-82.21}}} &
  {\color[HTML]{9a00d8} \textbf{-84.82}} &
  \textbf{-66.80} \\ \hline
\end{tabular}}
\label{tab:BD-rate}
\end{table*}

Overall, Bi-modal consistently outperforms the original KeyNode-driven method (TMM) across both complex and slow-motion sequences, demonstrating substantial compression gains while maintaining reconstruction quality. The improvements are most pronounced for sequences with complex, non-rigid motion, where Bi-modal achieves significant reductions in bitrate, from 26.04\% (\textit{Longdress - BiModal-A}) to 56.31\% (\textit{Levi - BiModal-M}). The performance gain on slow-motion sequences is smaller, as temporal redundancy in these sequences already favors accurate prediction.
Comparison with V-DMC further confirms that Bi-modal delivers higher coding efficiency across all sequences.

\subsubsection{Importance of Manual Segmentation}
The RD results demonstrate that automatic segmentation \textit{(BiModal-A)} performs nearly as effectively as manual segmentation \textit{(BiModal-M)} wherever available. For sequences such as \textit{Dancer}, where auto-segmentation is highly accurate, the R–D performance is virtually identical between the two modes. Even in challenging cases like \textit{Longdress}, where automatic segmentation errors are more pronounced, \textit{BiModal-A} still provides moderate improvements over the baselines. Notably, this indicates that high-quality compression performance can be reliably obtained automatically, which is critical for practical applications where manual annotation is costly or infeasible.

\subsection{Affine Combination Selection Insights}  
\subsubsection{Enhanced Prediction Quality}
\begin{figure}[h]
    \centering  
    \includegraphics[width=\linewidth]{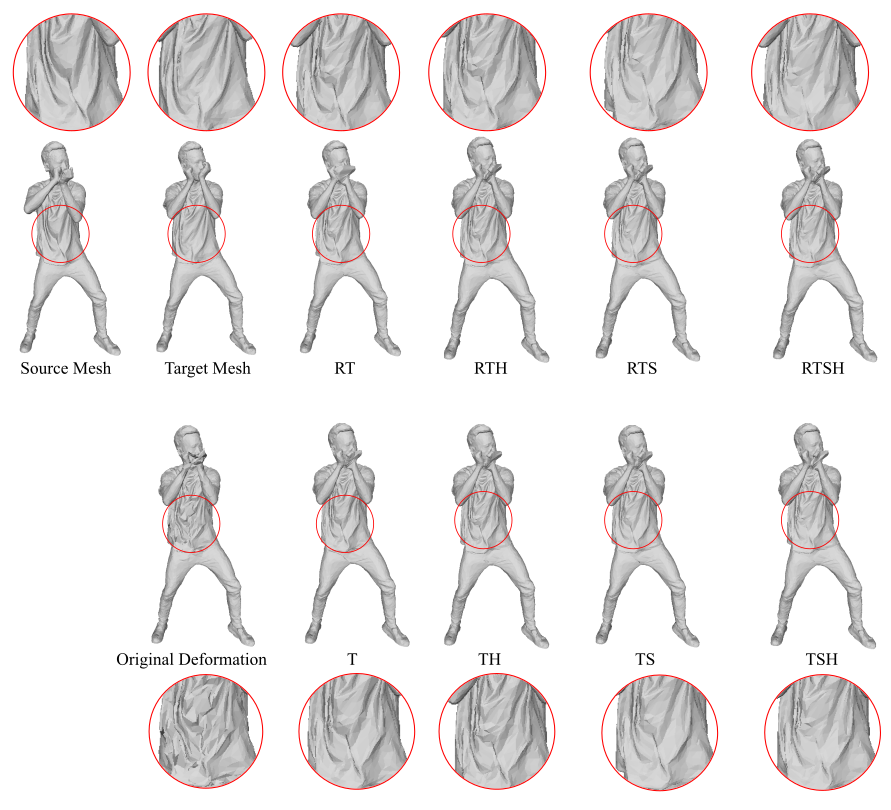}  
    \caption{Visualization of prediction quality for \textit{Dancer}'s frame 82 using Bi-modal's multiple affine combinations, compared with the original rigid prediction framework under the same set of key nodes. Source mesh: frame at the decoder, used to predict the next frame. Target mesh: the ground-truth next frame. }  
    \label{fig:affine_combinations_vis}  
\end{figure}

Here, we examine the role of affine combinations in enhancing prediction quality. Fig.~\ref{fig:affine_combinations_vis} illustrates the impact of different affine combinations used in Bi-modal’s prediction framework.  The figure highlights a scenario of rapid and highly non-rigid deformation, particularly in the unfolding of cloth. In the source mesh, the T-shirt contains multiple folds that must be smoothed out in the target frame. The original rigid-based deformation framework struggles to capture this fast, non-linear motion, resulting in noticeable artifacts. In contrast, Bi-modal’s affine combinations preserve the cloth's natural swirling motion. Certain combinations (e.g., RTSH and TSH) are especially effective in resolving fine-grained details such as the unfolding of folds, closely aligning with the target mesh.  These findings suggest that carefully chosen affine components are key to robust motion prediction in challenging deformation scenarios.  

\begin{figure}[h]
    \centering  
    \includegraphics[width=0.8\linewidth]{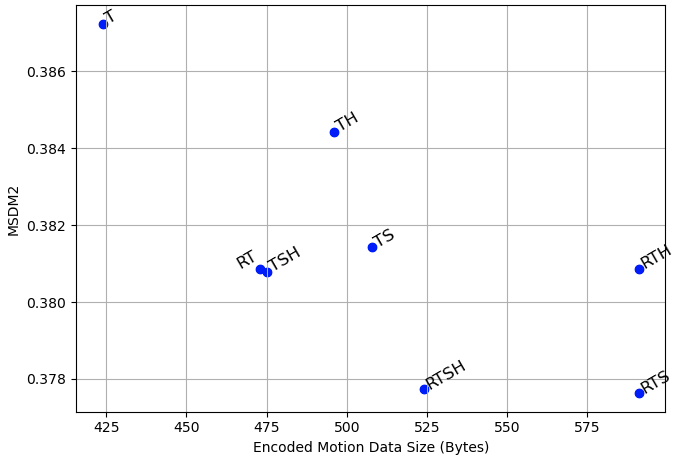}  
    \caption{Rate-distortion trade-offs of affine combinations. The rate and distortion are calculated after motion vector coding (without residual coding).}  
    \label{fig:affine_combination_tradeoffs}  
\end{figure}

While incorporating affine combinations improves prediction quality, it increases complexity and bitrate. 
Fig.~\ref{fig:affine_combination_tradeoffs} presents the R-D trade-offs for the different affine combinations of Fig.~\ref{fig:affine_combinations_vis}.  
The translation-only (T) model offers the lowest bitrate due to its minimal parameter count, yet it suffers from the highest distortion because of its limited capacity to represent complex motion. As additional components such as rotation (R), scaling (S), and shearing (H) are integrated, a substantial decrease in distortion is observed in models RT, TH, and TS, which more accurately capture the non-linear aspects of motion at the cost of a higher bitrate. The models TSH and RT strike a crucial balance, significantly reducing distortion while incurring only a moderate bitrate overhead compared to the simplest translation model.

While complex models (RTS, RTH, and RTSH) provide the lowest distortion, their superior performance is generally achieved at the highest bitrate. A notable exception is that for this specific frame, RTSH had lower bitrate than RTH or RTS, despite having more parameters. This counterintuitive result highlights the intricate relationship between parameter optimization and encoding efficiency.

Encoding motion parameters is not an independent operation for each component; rather, it is highly dependent on the values produced during the optimization phase. When a component is omitted from the model, the remaining components must compensate for its absence to minimize the prediction error. 
This compensation can force the values of the active parameters to be larger or to fall within a range that is statistically unfavorable for distribution-based compression algorithms. 
In contrast, a fully-featured model like RTSH allows the optimization algorithm to distribute the motion representation across all available parameters. This flexibility can lead to a more balanced and compressible set of values, ultimately resulting in a smaller encoded file size despite the higher parameter count.

These results highlight a key strength of the affine combination framework: \emph{flexibility}. By offering a spectrum of trade-offs between prediction quality, bitrate, and complexity, the method can be tailored to suit diverse application requirements. For instance, low-bitrate or real-time streaming scenarios may prioritize simpler combinations with reduced overhead, whereas offline high-fidelity reconstruction tasks can leverage richer transformations (e.g., RTSH) for maximum accuracy.

\subsubsection{Strong Robustness to Reduced Node Counts}
\label{sec:reduced_num_nodes}

\begin{figure}[h]
    \centering
    \includegraphics[width=0.75\linewidth]{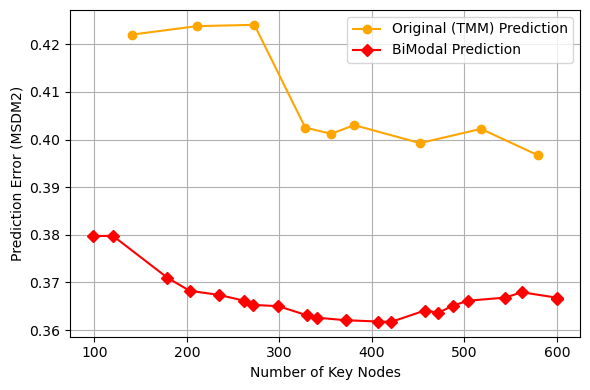}
    \caption{Prediction quality (prior to coding) versus the number of key nodes for frame 51 of the \textit{Dancer} sequence. }
    \label{fig:reduced_node_count}
\end{figure}

The proposed Bi-modal framework not only improves prediction accuracy but also demonstrates strong robustness when the number of key nodes is reduced. As shown in Fig.~\ref{fig:reduced_node_count}, Bi-modal consistently outperforms the baseline (TMM) across the entire range of node counts.  
For example, for \textit{Dancer}, Bi-modal maintains stable performance with about 150 key nodes, achieving lower distortion than TMM with about 300 nodes. Consequently, Bi-modal encodes all evaluated sequences using only half the number of key nodes used in TMM. 

\subsection{Ablation Studies}

\subsubsection{Impact of Segmentation-Guided Correspondence}

To quantify the contribution of the Segmentation-Guided Correspondence Estimation module, we conduct an ablation study by disabling this component and measuring the BD-rate (\%) relative to the complete Bi-modal pipeline, under the auto-segmentation setting. The results are summarized in Table~\ref{tab:seg-corr-off}.

\begin{table}[h]
\caption{BD-rate (\%) when removing the Segmentation-guided Correspondence Estimation from \textit{BiModal-A}. A negative BD-rate means turning off the component is better.}
\centering
\begin{tabular}{|c|c|c|c|}
\hline
\textbf{\begin{tabular}[c]{@{}c@{}}Complex-motion\\ Sequence\end{tabular}} &
  \textbf{BD-rate (\%)} &
  \textbf{\begin{tabular}[c]{@{}c@{}}Slow-motion\\ Sequence\end{tabular}} &
  \textbf{BD-rate (\%)} \\ \hline
Dancer             & 2.54  & Mitch                 & -4.11                 \\ \hline
Longdress          & -3.58 & Thomas                & -0.89                 \\ \hline
Levi               & 1.94  & Soldier               & -5.24                 \\ \hline
Basketball &  -1.52     & \multicolumn{1}{l|}{} & \multicolumn{1}{l|}{} \\ \hline
\end{tabular}
\label{tab:seg-corr-off}
\end{table}

The results reveal that the benefit of Segmentation-Guided Correspondence is strongly conditioned on segmentation accuracy and the presence of fast-moving segments. For fast-motion sequences such as \textit{Dancer} and \textit{Levi}, which exhibit acceptably accurate segmentation, the module achieves some quality gains when enabled. By contrast, in \textit{Longdress}, \textit{Basketball}, and \textit{Soldier}, segmentation is degraded by loose clothing covering the limbs, the interaction with an external object, or complex overlapping gear, making segmentation and its guided correspondence unreliable. In these cases, disabling the module actually improves performance.

\begin{figure}
    \centering
    \includegraphics[width=\linewidth]{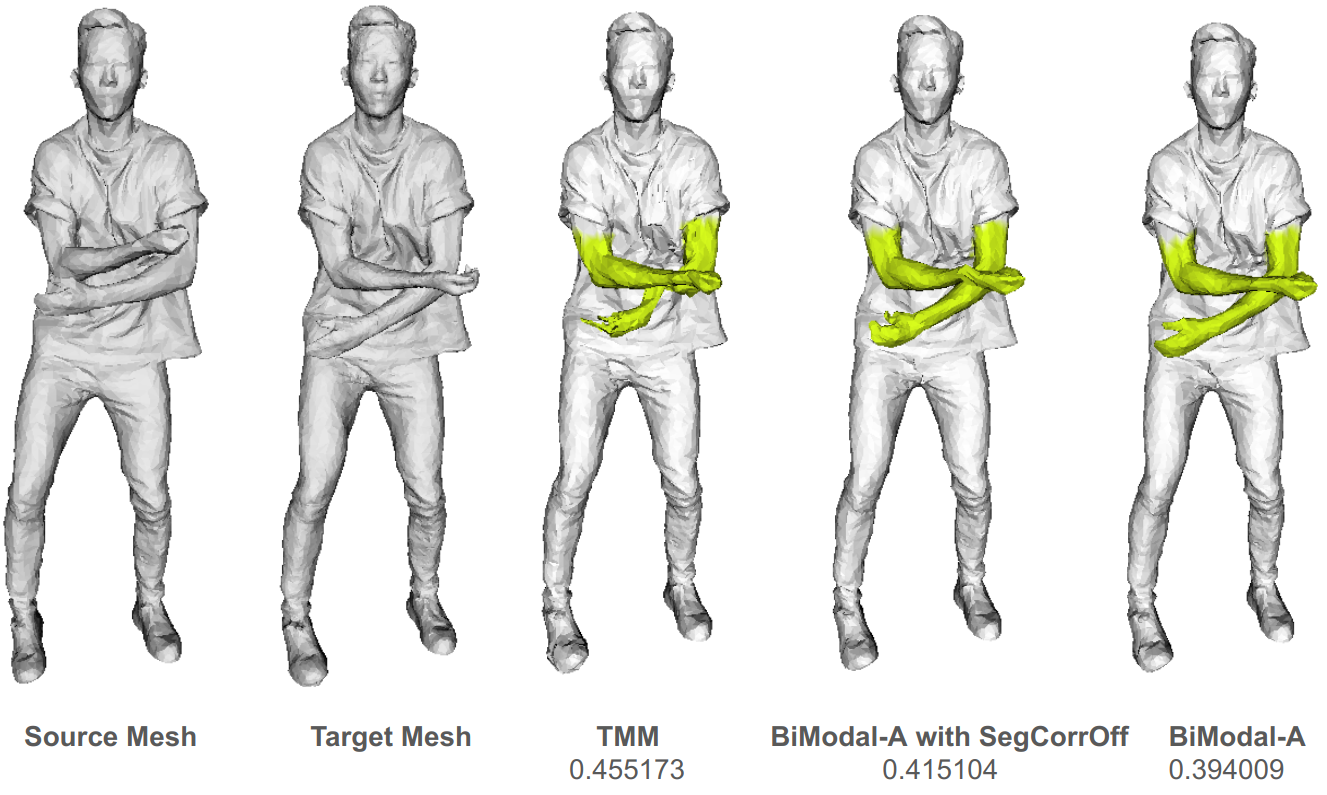}
    \caption{Prediction quality when the predicted motion is fast between frames (frame 201 to 202 in Dancer). The highlighted (yellow) areas are the places where the difference between methods is most noticeable. MSDM2 distortion is shown under each method.}
    \label{fig:seg_corr_off}
\end{figure}

The segmentation model we adopt \cite{Keito_3D} operates on a per-frame basis, without enforcing temporal consistency across frames. This lack of consistency makes cross-frame segment matching inherently difficult. Nevertheless, segmentation guidance remains valuable when certain body parts move too quickly for ICP-based deformation alone to track accurately. Fig.~\ref{fig:seg_corr_off} illustrates this for \textit{Dancer}: between frames 201 and 202, the rapid arm movement causes the baseline TMM to fail in predicting the motion. The ablation setting (\textit{BiModal-A with SegCorrOff}) provides a noticeable improvement over TMM, thanks to rigid/non-rigid area separation - which avoids the nearby nodes on the T-shirt or the opposite arm from interfering with the arm’s prediction, translation predictive coding, and the revised weighting scheme introduced in BiModal-A. However, it still struggles to capture the fast arm movement. When Segmentation-Guided Correspondence Estimation is enabled (BiModal-A), the prediction quality improves substantially, demonstrating the module’s advantage in fast-motion scenarios.

For slower-motion sequences such as \textit{Mitch}, \textit{Thomas}, and \textit{Soldier}, the benefits are far less pronounced. Here, the absence of rapid motion reduces the need for segmentation guidance, while inconsistencies at segment boundaries introduce errors that outweigh potential gains  (as shown in Table~\ref{tab:seg-corr-off}).

These findings highlight a nuanced role of segmentation-guided correspondence. The module proves especially effective in handling challenging regions such as fast-moving limbs, where ICP-based deformation struggles. However, its usefulness is limited by segmentation reliability. When segmentation is noisy due to clothing, occlusions, or irregular geometry, correspondence errors dominate. In practice, the decision to activate this module should depend jointly on segmentation confidence and the presence of fast body-part motion.

\subsubsection{Contribution of Translation Predictive Coding}

We further assess the role of the proposed Translation Predictive Coding by conducting an ablation study. Table~\ref{tab:T_pred_code_off} reports the performance when this module is replaced with the original Cauchy-based Huffman Encoding. Overall, Translation Predictive Coding improves compression efficiency, particularly for sequences with strongly correlated temporal motion. However, its benefits are not universal: when inter-frame correlations are weak, performance can deteriorate, as indicated by negative BD-rate values (e.g., \textit{Longdress}, \textit{Thomas}).  In our framework, Spatial Predictive Coding is applied only to the first P-frame, while subsequent P-frames in the GoF are encoded using Spatio-Temporal Predictive Coding. The observed results suggest that a more adaptive selection strategy may be required, one that dynamically determines the most suitable coding scheme for each P-frame based on motion characteristics.

\begin{table}[h]
\caption{BD-rate (\%) when replacing Translation Predictive Coding with the original Cauchy-based Huffman Encoding, relative to \textit{BiModal-A}. A negative BD-rate means removing the component is better.}
\centering
\begin{tabular}{|c|c|c|c|}
\hline
\textbf{\begin{tabular}[c]{@{}c@{}}Complex-motion\\ Sequence\end{tabular}} &
  \textbf{BD-rate (\%)} &
  \textbf{\begin{tabular}[c]{@{}c@{}}Slow-motion\\ Sequence\end{tabular}} &
  \textbf{BD-rate (\%)} \\ \hline
Dancer             & 8.23  & Mitch                 & 11.09                 \\ \hline
Longdress          & -0.83 & Thomas                & -9.78                 \\ \hline
Levi               & 0.5   & Soldier               & 2.77                  \\ \hline
Basketball         & 1.8      & \multicolumn{1}{l|}{} & \multicolumn{1}{l|}{} \\ \hline
\end{tabular}
\label{tab:T_pred_code_off}
\end{table}

\subsubsection{Significance of Weighting Schemes}

We examine the role of weighting schemes by comparing Bi-modal’s original design with two alternatives: the weights used in the original KeyNode-driven framework (TMM) and the scheme from~\cite{Chen2022}. Table~\ref{tab:weighting} reports the BD-rate when Bi-modal’s weights are replaced. The results show that Bi-modal’s weighting strategy is consistently more effective. Among the two, TMM weights provide closer performance to Bi-modal and remain relatively stable across sequences, whereas the scheme from~\cite{Chen2022} can fail dramatically in certain cases (e.g., \textit{Thomas}, with a BD-rate increase exceeding 90\%). The variation in performance also reveals content dependency: weighting has a stronger impact in complex, high-motion sequences (e.g., \textit{Dancer}, \textit{Levi}) than in low-motion ones (e.g., \textit{Soldier}). 

\begin{table}[h]
\centering
\caption{BD-rate (\%) when replacing Bi-modal's weighting scheme with others, relative to \textit{BiModal-A}. A negative BD-rate means replacing the component is better.}
\begin{tabular}{|c|c|c|}
\hline
\textbf{\begin{tabular}[c]{@{}c@{}}Sequence\end{tabular}} &
  \textbf{\begin{tabular}[c]{@{}c@{}}TMM Weights\\ BD-rate (\%)\end{tabular}} &
  \textbf{\begin{tabular}[c]{@{}c@{}}\cite{Chen2022}'s Weights\\ BD-rate (\%)\end{tabular}} \\ \hline
Dancer     & 18.46 & 30.29 \\ \hline
Longdress  & 12.53 & 16.92 \\ \hline
Levi       & 39.61 & 66.94 \\ \hline
Basketball & 13.14 & 30.58 \\ \hline
Mitch      & 24.79 & 71.88 \\ \hline
Thomas     & 16.60 & 91.17 \\ \hline
Soldier    & 6.69  & 23.81 \\ \hline
\end{tabular}
\label{tab:weighting}
\end{table}

\subsubsection{Role of Affine Combination Selection Schemes}
There are several strategies for selecting an appropriate affine combination to predict a given P-frame. One approach is to apply a certain affine decomposition selection across all P-frames in a GoF, optimizing the overall R-D trade-off via a Lagrangian formulation.
Alternatively, affine decomposition can be re-optimized at each P-frame independently. However, because the quality of each P-frame influences subsequent predictions, local optimization may fail to capture long-term temporal degradation effects. At the extreme, one could exhaustively search all possible affine decomposition combinations across P-frames, but the associated computational cost is prohibitive.  
In our Bi-modal codec, we adopt a more balanced strategy: the affine combination optimizer is applied only to the first P-frame of each GoF, as described in Section~\ref{sec:codec_architecture}. 

\begin{figure}[h]
    \centering
    \includegraphics[width=\linewidth]{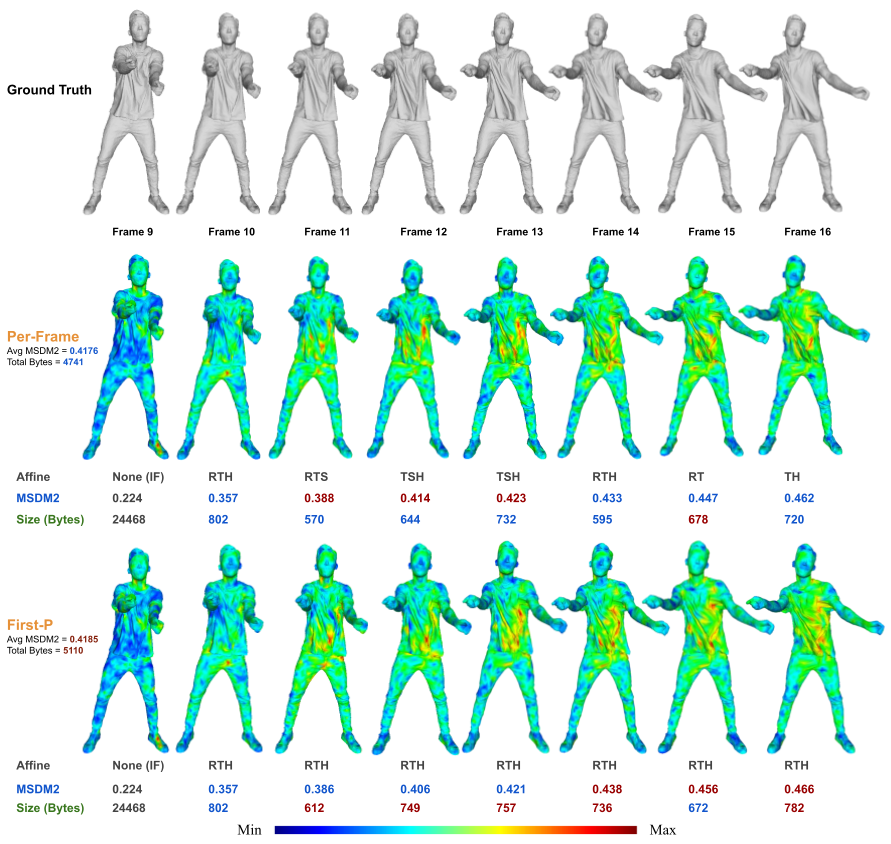}
    \caption{Decoded P-frames (colored by vertex-level MSDM2 errors) under two affine combination optimization schemes in a GoF of 8 frames from \textit{Dancer}. Average distortion and total bitrate are reported for all P-frames.}
    \label{fig:oracle_vs_fixed}
\end{figure}

In this section, we compare different optimization schemes, including our \textit{First-P} approach and a \textit{Per-Frame} scheme that re-optimizes every P-frame.
Due to the high computational cost of the \textit{Per-Frame} approach, we restrict our evaluation to 8 frames of \textit{Dancer}, with decoded mesh visualizations shown in Fig.~\ref{fig:oracle_vs_fixed}. 

For the first P-frame of the GoF (frame 10), both \textit{First-P} and \textit{Per-Frame} select the same affine combination, since both run the optimizer at this point. From the second P-frame onward, the strategies diverge. \textit{Per-Frame} dynamically adapts the combination (e.g., selecting RTS) to achieve lower bitrate at the cost of slightly higher distortion, while \textit{First-P} continues to reuse the combination fixed at frame 10 (RTH), yielding higher bitrate but lower distortion in the short term.  

This trade-off shifts as motion evolves: by later P-frames, the fixed RTH combination in \textit{First-P} no longer represents the motion well, leading to both higher bitrate and higher distortion compared to the updated choices in \textit{Per-Frame}. When averaged across all P-frames, \textit{Per-Frame} achieves both lower distortion and lower bitrate. 
The drawback, however, is that its complexity scales linearly with the number of P-frames, making it substantially more complex in practice.

\section{Conclusion and Future Work} 

In this paper, we introduced a Bi-modal Prediction Method that significantly advances dynamic mesh compression by allowing a KeyNode-driven codec to efficiently handle complex, non-rigid deformations. Our method integrates semantic segmentation to selectively apply a more expressive affine transformation model to deformation-rich regions (e.g., loose, fast-moving clothing), while retaining simpler, rigid models for other areas. This adaptive strategy, guided by a Lagrangian R-D optimization, ensures a precise balance between motion fidelity and coding efficiency.

Our evaluations demonstrate that this approach achieves superior R-D performance, particularly for sequences with complex motion. Our method consistently outperforms the baseline KeyNode-driven codec \cite{Hoang_2025_KeyNode} by a substantial margin, with an average BD-rate saving of 33.81\%. A key insight from our analysis is the power of the affine decomposition; by factorizing transformations into their fundamental components—rotation, translation, scaling, and shearing—and encoding only what is necessary, we achieve substantial improvements in prediction quality with minimal bitrate overhead. 
Our ablation studies provided key insights into the architectural design, revealing the advantage of the weighting scheme, the data-dependent effectiveness of our predictive coding for translations, the segmentation-quality-dependent role of segmentation-guided correspondence, and the trade-offs involved in selecting affine combinations. In essence, this work presents a robust, scalable, and practical solution to efficiently compress intricate real-world human motion data.

Our work opens several promising avenues for future research. A primary area for improvement is the integration of more sophisticated segmentation and temporal consistency modules. While our current method is robust to segmentation inaccuracies, a more advanced segmentation network that can enforce temporal consistency across frames would improve the effectiveness of the segmentation-guided correspondence module, especially in challenging sequences such as \textit{Longdress}. 

Another key area is the development of a more adaptive coding selection strategy. Our ablation study on Translation Predictive Coding revealed that its effectiveness is context-dependent. Future work could explore a dynamic, content-aware selection scheme that decides on a per-frame basis whether to use spatial, spatio-temporal, or no prediction for translation vectors, based on local motion characteristics and an R-D metric. This would further optimize compression efficiency across a wider range of motion types.

Another promising direction is to explore a more sophisticated, adaptive selection method for affine combinations. Instead of fixing the affine combination for an entire GoF, future work could explore a look-ahead optimization strategy. This method would not only consider the current P-frame's distortion and rate but also estimate the long-term impact on subsequent frames, to minimize the accumulation of prediction errors. This approach would require a model to predict future frame complexity and motion characteristics to make informed, forward-looking decisions.

Finally, there is significant potential in exploring hybrid coding approaches that combine our explicit, KeyNode-driven framework with emerging implicit neural representations. Instead of relying on a pre-defined set of affine components, a learned neural model could predict and encode deformations, potentially capturing even more complex non-linear motion with greater accuracy. This would require solving new challenges related to bitrate control and ensuring interpretability in a neural context. For instance, one could train a neural network to output the optimal affine decomposition and quantization parameters for each key node, creating a data-driven approach to our existing framework.

\section*{Acknowledgement}
We thank Andrew J.\ Chen, a high school student at Canyon Crest Academy, for contributing to the segmentation of the \textit{Dancer} and \textit{Thomas} sequences.

\bibliographystyle{IEEEtran}
\bibliography{egbib}

\vfill

\end{document}